\documentclass[preprint,times]{aastex} 
\usepackage{emulateapj5}  
\usepackage{psfig}

\newcommand{\astroph}[1]{{astro-ph/#1}}
\newcommand{\sci}{{Science }}

\begin{document}
\submitted{Accepted for publication in the Astrophysical Journal, Letters}

\title{Signatures in a Giant Radio Galaxy of a Cosmological Shock Wave at
Intersecting Filaments of Galaxies}

\author{Torsten A. En{\ss}lin\altaffilmark{1,2,3}, Patrick Simon\altaffilmark{3},
Peter L. Biermann\altaffilmark{3},\\  Ulrich Klein\altaffilmark{4}, Sven Kohle\altaffilmark{4},
Philipp P. Kronberg\altaffilmark{2}, Karl-Heinz
Mack\altaffilmark{5,4}}

\affil{$^1$ Max-Planck-Institut f{\"u}r Astro\-phy\-sik,
Karl-Schwarz\-schild\--Str. 1, D-85741 Gar\-ching, Germany}


\affil{$^2$ Department of
Physics, University of Toronto, 60 St. George Street, Toronto, M5S1A7,
Canada}

\affil{$^3$ Max-Planck-Institut f{\"u}r Ra\-dio\-astro\-no\-mie, Auf dem
H{\"u}gel 69, D-53010 Bonn, Germany}

\affil{$^4$ Radioastronomisches Institut,
Universit{\"a}t Bonn, Auf dem H{\"u}gel 70, D-53121 Bonn,
Germany}

\affil{$^5$ Istituto di Radioastronomia del CNR, Via P. Gobetti, 101
, I-40129 Bologna, Italy}

\begin{abstract}
Sensitive images of low-level, Mpc-sized radio cocoons offer new
opportunities to probe large scale intergalactic gas flows outside
clusters of galaxies.  New radio images of high surface brightness
sensitivity at strategically chosen wavelengths of the giant radio
galaxy NGC~315 \citep{mack97AAS,mack98AA} reveal significant
asymmetries and particularities in the morphology, radio spectrum and
polarization of the ejected radio plasma.  We {argue that} the
combination of these signatures provides a sensitive probe of an
environmental shock wave. Analysis of optical redshifts in NGC~315
vicinity confirms its location to be near, or at a site of large-scale
flow collisions in the 100 Mpc sized Pisces-Perseus Supercluster
region. NGC~315 resides at the intersection of several galaxy
filaments, and its radio plasma serves there as a `weather station'
\citep{1998Sci...280..400B} probing the flow of the elusive and
previously invisible IGM gas.  If our interpretation is correct, this
is the first indication for a shock wave in flows caused by the
cosmological large scale structure formation, which is located in a
filament of galaxies.  The possibility that the putative shock wave is
a source of gamma-rays and ultra high energy cosmic rays is briefly
discussed.
\end{abstract}
\keywords{acceleration of particles -- shock waves -- intergalactic
medium -- large-scale structure of universe -- galaxies: individual
(NGC 315) -- galaxies: clusters: individual
(Pisces-Perseus-Supercluster)}

\section{Introduction}
Beginning with the first detailed images, it was speculated that
the Z-like morphology of the giant, Mpc sized radio galaxy NGC~315
(Fig. \ref{fig:ngc315spix}) is affected by environmental influences
\citep{bridle76}. {New high frequency maps unveil a unusually flat
radio spectrum at the end of the Western (W) lobe, indicating that the
lobe's CR electrons are freshly re-energized far from AGN or first
termination shock. No comparable structure is found in the Eastern (E)
lobe.}  At NGC~315's location several sub-filaments of the
Pisces-Perseus Supercluster intersect to form the beginning of its
main filament (Fig. \ref{fig:ppsc}). Gravitationally driven,
supersonic flows of intergalactic gas and galaxies onto and along
these filaments, and towards the clusters of galaxies will inevitably
collide, and form large scale shock waves on the largest conceivable
scales \citep{1998ApJ...502..518Q,Miniatietal2000a}. Therefore
NGC~315 resides in an environment where shock waves are indeed
expected to be present. In this paper we investigate the hypothesis
that such a wave re-energizes the W lobe's electron population.

Large-scale shock waves are the primary heat source in the wider
intergalactic medium (IGM), and they also allow measurement of its
unknown entropic history \citep{tozzi2000}. They have been proposed
as possible generation sites of intergalactic magnetic fields
\citep{1997ApJ...480..481K,1998A&A...335...19R}, of ultra high energy
cosmic rays \citep{Norman95,kang96,1997MNRAS.286..257K,so00}, and of
the diffuse gamma ray background \citep{Loeb2000,totani2000}.  Prior
to this work, merger and accretion shock waves have been detected only
inside of galaxy clusters by temperature structures in the X-ray
observable cluster gas \citep{1993ApJ...419...66M,1998ApJ...500..138D}
or by radio emission from shock accelerated electrons
\citep{ensslin1998,1999ApJ...518..603R,ensslin2000}.  Here, the
{putative} shock wave is located in a galaxy filament, on a scale
that is 10 - 100 times the dimension of a cluster and with a gas
density that is orders of magnitude {lower}. The new data therefore
{raises a new possibility of} the precise detection and probing of
flows in the ``filament and bubble'' morphology predicted by the
theory of cosmological large scale structure formation
\citep[cf.]{Bond96}.

\section{NGC 315: a cosmic weather station}
The giant radio galaxy NGC~315 has two straight, bright radio jets,
which form termination shock waves at a distance of 860 kpc (E) and
420 kpc (W) from the galaxy where they are stopped by the
intergalactic medium (assuming $H_0 = 65\,{\rm km\,s^{-1}\,Mpc^{-1}}$
and an EdS cosmology). In the standard model of extended radio sources
these shock waves serve as accelerators for the radio emitting
relativistic electrons of the jet material.  After shock passage the
electron energy ($E$) distribution function is given by a power-law
$f(E)\,dE \sim E^{-s}\,dE$ within the radio-observable energy
range. This leads to synchrotron emission with a power-law radio
frequency ($\nu$) spectrum $F_\nu \sim \nu^{-\alpha}$, with\\[-2em] 
%
\psfig{figure=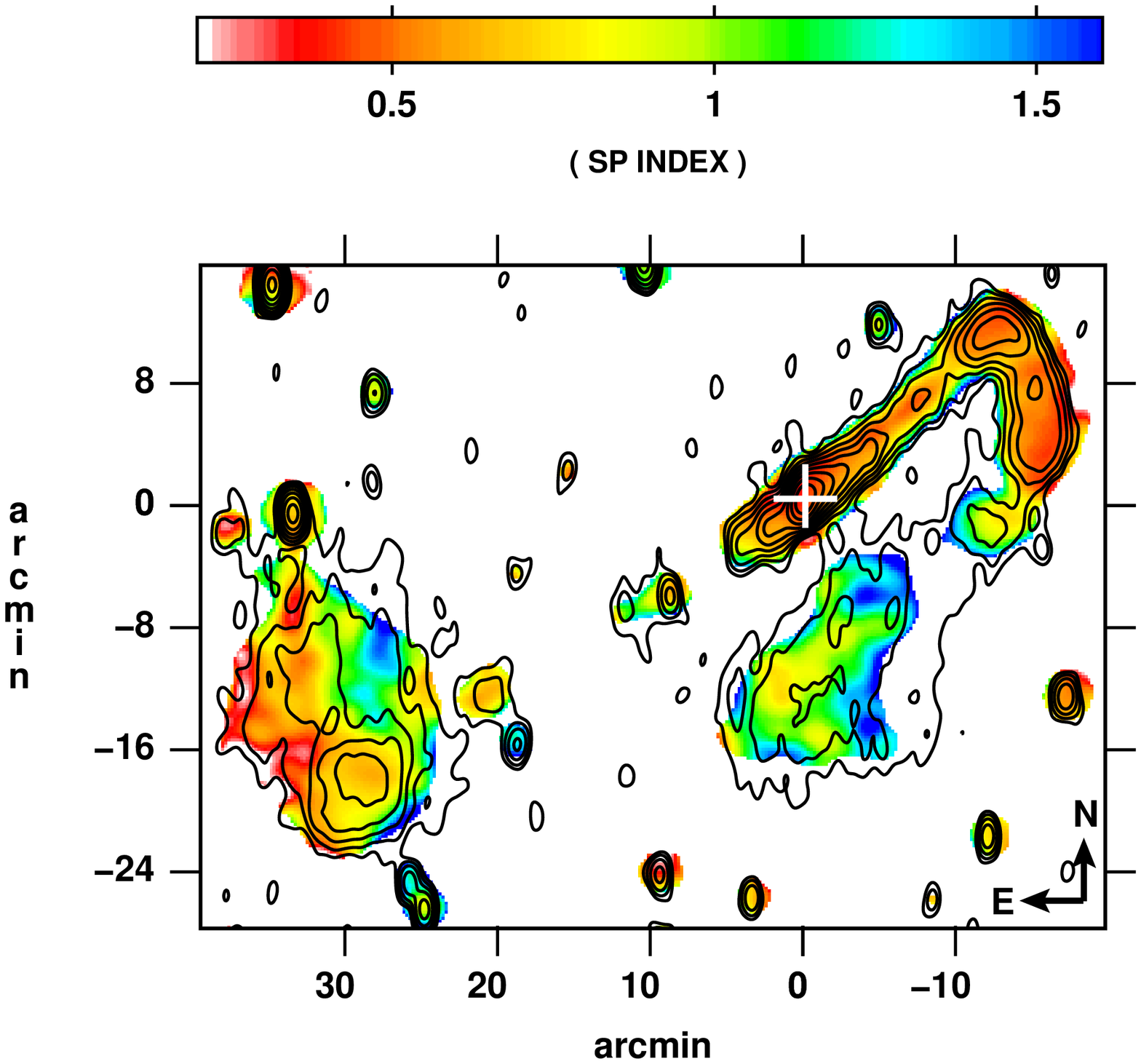,width=0.5\textwidth,angle=0}
\figcaption[ensslin_fig1.ps]{\label{fig:ngc315spix} 327 MHz radio
map of NGC~315 (contours) superimposed on spectral index map between
327 MHz and 600 MHz \citep{mack98AA}. The position of the galaxy is
marked.  Note that the spectral index in the W tail might be flatter
than displayed due to possibly missing 600 MHz flux of extended structures.\\[1em]}
%
\noindent $\alpha = (s-1)/2$.  {In the course of the outflow,}
subsequent radiative synchrotron and inverse Compton cooling leads to
a steepening of the electron spectrum at the highest
energies. Together with adiabatic expansion losses this {can be
expected to produce a progressive} fading of the radio plasma
beginning at the highest frequencies. When the fading reaches the
lowest observable frequencies the radio plasma {becomes} undetectable
and can be called a `fossil'.

Precession of the jet axis of NGC~315 appears to have produced radio
plasma trails of the termination shocks of the jets \citep[see our
Figs. \ref{fig:ngc315spix} and \ref{fig:ngc315}]{bridle76}. The
observable part of the E trail is short at higher frequencies; 130 kpc
at 1.4 GHz, but 480 kpc at 327 MHz. This conforms to the typical
behavior of cooling radio trails. However, an unexpected result was
that the W trail could be detected even at the {relatively} high
frequency of 4.8 GHz over a distance of 600 kpc from the termination
shock of the W jet, {and up to 800 kpc at 327 MHz}.  The spectral
index measured between 327 and 600 MHz is $\alpha_{327}^{600} \approx
0.75 ... 1.25$ -- {quite flat}, as can be seen in
Fig. \ref{fig:ngc315spix}. It remains flat even at higher frequencies
up to 10.6 MHz \citep{mack98AA}.  The morphologies of the two trails
are also different, in that the E trail is roughly as one expects for
a precessing jet, but the W trail shows a bending 200 kpc away from
the hot spot. Finally the W trail has an average radio polarization of
$32 \pm 9$ \% at 10.6 MHz.

Three properties of the W trail suggest that an originally symmetrical
radio galaxy is ``falling'', together with its surrounding IGM
environment, into an intergalactic shock wave. These are: (i) the
unusually flat spectral index up to high radio frequencies at large
distances from the point of ejection, (ii) the radio polarization
characteristics, and (iii) the bending. The most {striking} fact (i)
is that the electron population {is being} re-energized in the
fossil radio plasma on a large scale. {This region is} remote from
any possible galactic source of fresh electrons. And it is
separated from the jet termination point by a 
\noindent steep spectrum region.
%
\psfig{figure=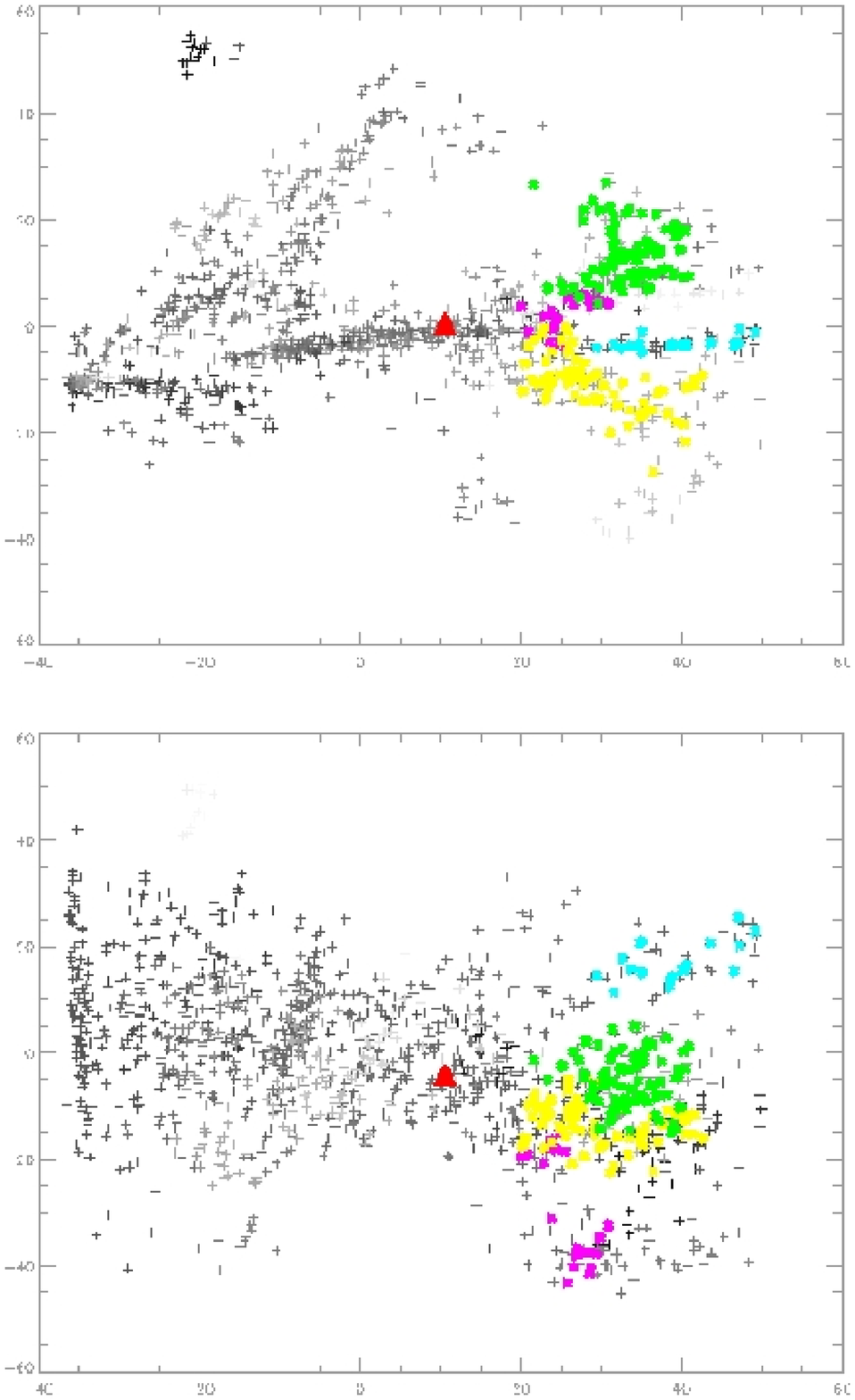,width=0.55\textwidth,angle=0}
\figcaption[ensslin_fig2.small.ps]{\label{fig:ppsc} Galaxy distribution in
the Pisces-Perseus Supercluster sky projection (top) and seen from
above in redshift space (bottom). The third dimension is encoded in
the grey-scale. Distances are given in Mpc. A density filter was
applied to enhance the visual significance of the filamentary
structure.  The galaxies of the intersecting filaments are colored
(blue = II, pink = III, yellow = IV, green = V).  The estimated
velocity dispersions of the different filaments I-V are $\sigma =
399\pm 14, 169\pm 22, 92\pm 26, 221 \pm 26, 171\pm 33 {\rm
km/s}$. The individual filaments cannot be separated in the
intersection region due to their overlap in the redshift space which
is caused by their velocity dispersions. Therefore it is not clear
which of them hosts NGC~315 (red triangle).\\[1em]}
%

A natural particle acceleration mechanism is compression due to a
large scale shock. In the following we pursue this model as the
supposed cause of the spectral and polarization properties of the
western lobe of NGC315. The consequence is a strong increase of the
radio emissivity, which is also expected to be coherently polarized,
due to the large scale alignment of the compressed magnetic fields
(ii). The degree of polarization depends largely on the angle between
the line of sight and the shock normal. The observed polarization
suggests this angle to be $5^\circ ...20^\circ$
\citep{ensslin1998}. However high polarization, also seen in many
other extended radio sources, is not a unique test of our putative
large sale shock scenario.  The shock may well also explain the
bending of the W trail (iii) (Fig. \ref{fig:ngc315}).

Alternative explanations of the bending are an environmental
subsonic wind, or a projection of a special precession history of the
jets. However precession should have produced a comparable structure
in the eastern lobe and a possible wind cannot explain the spectral
flattening at the lobes end.

\section{Pisces-Perseus Supercluster of Galaxies}

In order to estimate the expected shock strength we analyzed the
redshift distribution of galaxies in the Pisces-Perseus Supercluster
using the CfA2 redshift catalog \citep{1990ApJS...72..433H}.  Since we
are interested in the galaxies' velocity dispersion $\sigma^2 =
\langle \vec{v}^2 \rangle$ we had to disentangle the positional
information in the redshift by means of statistics.  We identified
filaments by eye and estimated their orientations in redshift space
using the main axis of the inertia tensor of the galaxy distribution
within the chosen sub-volumes (Fig. \ref{fig:ppsc}).  These will
approximately be identical to the real-space axes if we can assume
that the (statistical) distribution of galaxies in a filament has a
cylindrical symmetry with respect to the filament axis, and that the
velocity dispersion of galaxies is isotropic. These assumptions allow
us further to subtract the angular diameter (defined as the angular
dispersion times the distance) of the filament from the redshift
dispersion, yielding a rough estimate of the required velocity
dispersion.

The main filament (I) has $\sigma_{\rm I} = 400\pm 15 \,{\rm km/s}$
whereas the smaller filaments (II...V) have $\sigma_{\rm II...V}$
between $90...220\,{\rm km/s}$, a substantial difference. Since the
velocity dispersion of the galaxies and the IGM-gas of the filaments
should be comparable, this translates into temperatures of $kT_{\rm
I}= 280\,{\rm eV}$ and $kT_{\rm II...V} = 15 ... 85\,{\rm eV}$. The
gas in one of the smaller filaments might get heated by a shock wave
when it flows into the deeper gravitational potential of the main
filament. Such a shock wave is expected to have a temperature jump
$kT_2/kT_1 = kT_{\rm I}/kT_{\rm II...IV} = 3.3 ... 20$, a compression
ratio $R_{\rm gas} = 2.9...3.8$, a pressure jump $P_2/P_1 =
9.6... 75$, and a shock velocity of $v_{\rm shock} = 460...520\,{\rm
km/s}$ \citep{landau}. The subscript $1$ ($2$) refers to the upstream
(downstream) region of the shock wave. In all following estimates,
the indicated parameter ranges result from the above shock strength
ranges.

\section{Particle Acceleration}
Quantities such as the shock wave orientation, projection geometry,
speed of the jet precession, and galaxy motion cannot be completely
constrained by the available data. We restrict ourselves only to the
assumption that the W trail became compressed several 10 Myr after
deposition of the radio plasma in {a shock caused by one of the
merging galaxy filaments}.  {A} shock wave is only able to penetrate
into {the} radio plasma if the (unknown) internal sound velocity is
smaller than the shock speed.  If it is, {diffusive shock}
acceleration of the existing relativistic electron population at {the}
shock would explain the re-population of the higher electron
energies. If not, the alternative possibility is that the radio plasma
gets adiabatically compressed by the shock. {This} also leads to an
energy gain of the relativistic electron population. We demonstrate
{below} that either scenario is able to explain the unusual W trail of
NGC~315.
%
\psfig{figure=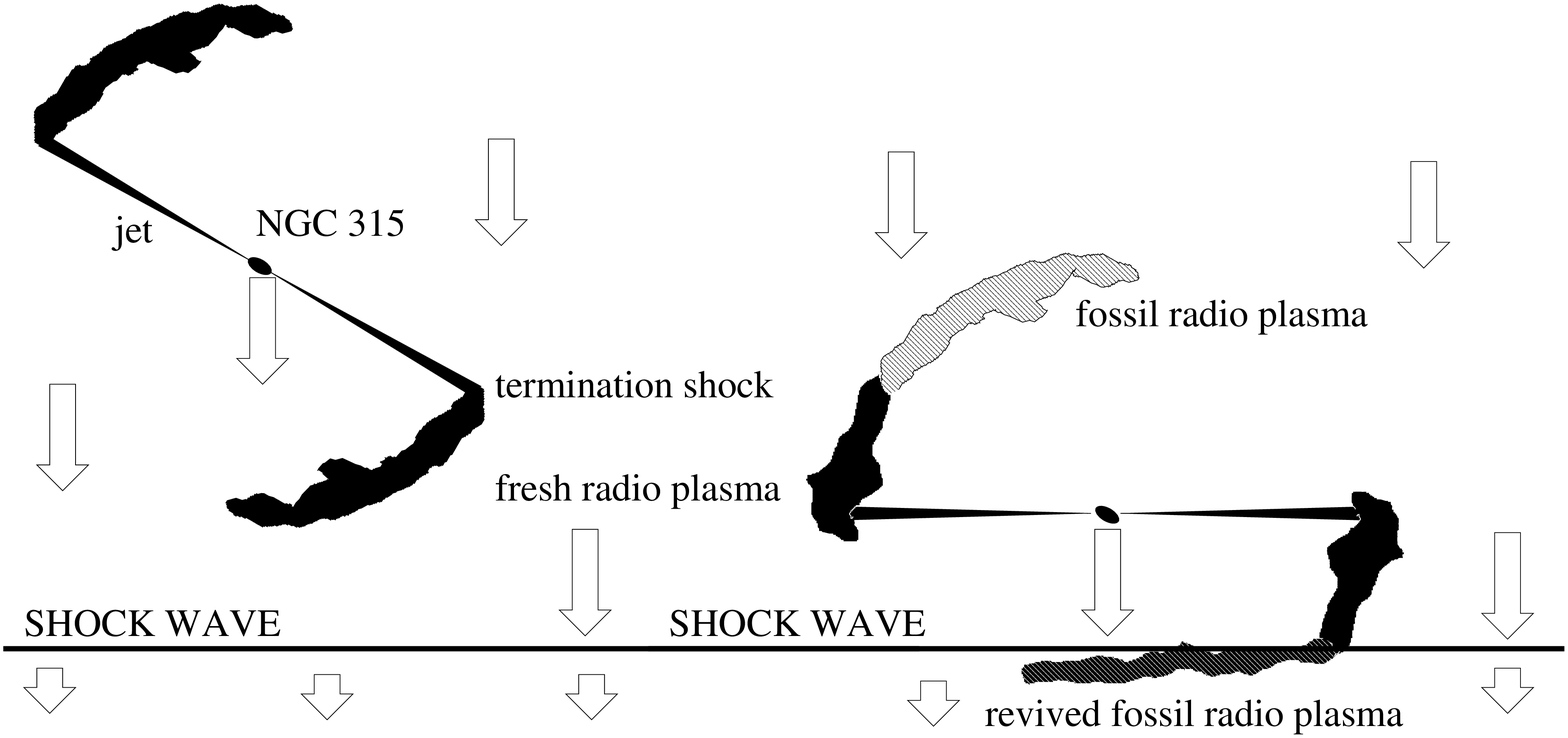,width=0.45\textwidth,angle=0}
\figcaption[ensslin_fig3.ps]{\label{fig:ngc315} Sketch of the
discussed scenario for NGC 315 and its environment. Left: the
undisturbed radio galaxy approaching a shock wave of the embedding gas
flow. Right: the observed present stage, where the preceding radio
plasma gets revived in the shock wave.}

\subsection{Shock Acceleration} Shock acceleration 
produces a power law electron {energy} distribution with a spectral
index $s = (R+ 2)/(R-1)$ \citep{drury83}, where $R$ is the compression
ratio of the accelerating shock wave inside the radio plasma. Since
the radio plasma can have a relativistic equation of state, $R$ can in
principle be bigger than $R_{\rm gas}$. But using $R=R_{\rm gas}$ for
illustration gives $s = 2.1...2.6$, which leads to a radio spectral
index of $\alpha = 0.53...0.79$. Superposition of re-accelerated
electrons with different ages produces a steepening of $\alpha$ by 0.5
above some break frequency. Cooling after the re-acceleration
successively removes the highest frequency emission. The observed
spectral indices (Fig. \ref{fig:ngc315spix}) are consistent with the
expectations in the shock acceleration scenario.
A key physical test is whether the shock has enough power to generate
re-accelerated electrons in a sufficient number. The rate of kinetic
energy dissipation by the shock wave in the surface $A = 200 \times
600 \, {\rm kpc^2}$ of the `revived' W trail is $Q =
\frac{1}{2}\,\varrho \, v_{\rm shock}^3\,A\, (1-1/R_{\rm gas}^2) = 2.5
... 3.8 \cdot 10^{42}\,{\rm erg/s}$. Here $\varrho$ is the upstream
mass density which we assume to be $\varrho = 5\cdot 10^{-29}\,{\rm
g\, cm^{-3}}$, following the estimate in \cite{mack98AA} of a
few times $1.7 \cdot 10^{-29}\,{\rm g\, cm^{-3}}$.  The radio
luminosity of the revived W trail is only $3.7\cdot 10^{40}\,{\rm
erg/s}$ \citep{mack98AA}, $\le 1\%$ of the dissipated kinetic energy
$Q$. This indicates that the radio electron population can be
explained with a reasonable shock efficiency.

\subsection{Adiabatic Acceleration} Adiabatic compression increases the
energy density of the relativistic electrons and magnetic fields by
$R^{4/3}$. {Here,} R is the compression of the radio plasma in the
shock, given for a fully relativistic equation of state by $R =
(P_2/P_1)^{3/4} = 5.4...26$.  The synchrotron luminosity for an
electron population with spectral index $s$ increases by $R^{2\,
s/3}$.  The surface brightness increases probably less drastically,
due to the more extended morphology of the radio plasma after
compression.
An observable brightening is only possible if the upper electron
energy cutoff $E_{\rm max}$ is high enough to allow synchrotron
radiation at the observed frequencies.  Synchrotron and inverse
Compton aging of the fossil electron population should have produced a
cutoff at $E_{\rm max} = 310 \,{\rm GeV} \,(t/{\rm Myr})^{-1}\,((u_B +
u_{\rm cmb})/({\rm eV\,cm^{-3}}))^{-1}$ \citep{ensslin2000}, where $t$
is the age of the fossil radio plasma (after passage of the
termination shock), $u_B$ the magnetic and $u_{\rm cmb} = 0.28\, {\rm
eV\, cm^{-3}}$ the cosmic microwave background energy density. $E_{\rm
max}$ corresponds to a cutoff in the synchrotron spectrum at $\nu_{\rm
max} = 16 \,{\rm MHz}\,(E_{\rm max}/{\rm GeV})^2\,(B/{\mu{\rm G}})$.
The compression increases $E_{\rm max}$ by $R^{1/3}$, the magnetic
field strength $B$ by $R^{2/3}$, and thus the upper frequency cutoff
by $R^{4/3}$.  This can be used to infer a maximum age the fossil
radio plasma can have, for the {compression} wave to revive the
emission at frequency $\nu$, which is
\begin{eqnarray}
t_{\rm max} &=& \frac{39 \, {\rm Myr} \,R^{2/3}}{(u_B + u_{\rm cmb})/({\rm
 eV\,cm^{-3}})}\, \sqrt{ \frac{B/{\rm \mu
 G}}{\nu/{\rm GHz}}} \,.\nonumber
\end{eqnarray}
Given that detection was made at a frequency as high as $\nu = 10.6$
GHz, and the assumption of an initial magnetic field strength of $B =
0.5\,\mu$G \citep{mack98AA} in the trail we obtain $t_{\rm max} = 93
... 260$ Myr.  More detailed calculations, following the formalism
described in \cite{ensslin2000}, show that an age of $< 100$ Myr is
more realistic when we allow for the enhanced cooling during the
process of compression.

\section{Gamma rays}
It has been proposed that shocks of the large scale structure
formation accelerate electrons up to energies at which they can
scatter the cosmic microwave background into the gamma-ray regime
\citep{Loeb2000,totani2000}. If these electrons receive 5 \% of the
dissipated energy and have a flat spectral index of $s = 2$ they could
explain the observed diffuse gamma ray background. The {putative}
shock wave {indicated} by NGC~315 {would imply that we have} an
ideal testbed for this hypothesis. We assume the same efficiency, but
allow for the steeper spectral index due to non-maximal compression
($R_{\rm gas} = 2.9...3.8 < 4 \rightarrow s = 2.1...2.6 >2$). This
gives an expected gamma ray luminosity of
$ F_\gamma(>E_\gamma) = 3\cdot~10^{-9}\,(E_\gamma/{\rm 100\,
MeV})^{-1.03} \,{\rm photon \,\,cm^{-2}\,s^{-1}}$ for $s=2.1$ and
$F_\gamma(>E_\gamma) = 10^{-11}\,(E_\gamma/{\rm 100\,
MeV})^{-1.29}\,{\rm photon \,\,cm^{-2}\,s^{-1}}$ for $s=2.6$, where
the shock surface was assumed to be 10~Mpc$^2$. This is consistent
with upper limits from the EGRET satellite, but it could be detectable
by the future GLAST experiment.

\section{Conclusion}
We have shown that the properties of the ejected radio plasma from the
giant radio galaxy NGC~315 can be naturally explained by a transit
through a powerful shock wave. {Such a} shock wave {should result}
from collisions of cosmological flows along the intersecting galaxy
filaments in the Pisces-Perseus Supercluster.  This kind of
observation provides a new method of exploring the tenuous baryonic
matter between galaxies
\citep{ensslin1998,mack98AA,1997ApJ...480...96W}. The {proposed
scenario} for NGC~315 {suggests} that the IGM already becomes
shocked outside of clusters \citep{tozzi2000}, and that the proposed
large-scale structure shock wave origin of the gamma ray background
should be testable \citep{Loeb2000,totani2000}.  {Of course an
independent confirmation of the shock is still necessary. Possible
strategies are very deep X-ray observations and the search for low
frequency, diffuse radio sources associated with the shock -- similar
to the cluster radio relic sources found at cluster merger shock
waves. The ability of radio galaxies to trace environmental shock waves
would be further supported if other examples of interaction of radio
sources with shock wave could be identified. Also, reactions of the
interstellar medium of galaxies which are passing through a shock wave
-- as in Abell~1367
\citep{1995A&A...304..325G,ensslin1998,1998ApJ...500..138D} -- might be
used to reveal shock waves outside clusters of galaxies.}

Finally we note that one of the three doublets of ultra-high energy
cosmic ray particle arrivals reported by the AGASA experiment
\citep{hayashida96} lies in the direction of the Pisces-Perseus
Supercluster. This may be an indication of a connection between
large-scale intergalactic shocks and the acceleration of cosmic rays
\citep{Norman95,kang96,1997MNRAS.286..257K,so00}.

\acknowledgments 
Parts of this research were supported by the {National Science and
Engineering Research Council of Canada} (NSERC), and by the European
Commission, TMR Programme, Research Network Contract ERBFMRXCT96-0034
``CERES''.

\end{document}